\documentclass[12pt]{article}

\usepackage{latexsym,amsmath,amssymb}
\usepackage{graphicx}

\evensidemargin \oddsidemargin

\begin{document}


\begin{titlepage}

\renewcommand{\thefootnote}{\fnsymbol{footnote}}



\vspace{15mm}
\baselineskip 9mm
\begin{center}
  {\Large \bf Schwarzschild-de Sitter black hole \\ from entropic viewpoint}
\end{center}

\baselineskip 6mm
\vspace{10mm}
\begin{center}
  Ee Chang-Young${}^{a}$\footnote{Email: cylee@sejong.ac.kr},
  Myungseok Eune${}^{b}$\footnote{Email: younms@sogang.ac.kr},
  Kyoungtae Kimm${}^{a}$\footnote{Email: helloktk@naver.com}, and
  Daeho Lee${}^{a}$\footnote{Email: dhleep@gmail.com}\\
  \vspace{3mm}
  {\sl ${}^a$Department of Physics and Institute of Fundamental Physics,\\
    Sejong University, Seoul 143-747, Korea}
  \\
  \vspace{3mm}
  {\sl ${}^b$Basic Science Research Institute, Sogang University, Seoul
    121-742, Korea}

\end{center}

\thispagestyle{empty}

\vfill
\begin{center}
{\bf Abstract}
\end{center}
\noindent
In a Schwarzschild-de Sitter space, we consider an equipotential
surface which consists of two holographic
screens.  
Adapting the Bousso-Hawking's reference point of vanishing force, we
divide the space into two regions, which are from the reference point
to each holographic screen.  These two regions can be treated as
independent thermodynamical systems, because the Bousso-Hawking
reference point with zero temperature behaves like a thermally
insulating wall. The entropy obtained in this way agrees with the
conventional results; i) when the holographic screens lie at the black
hole and cosmological horizons, ii) in the Nariai limit.
\\ [5mm]
\vspace*{1cm}

\vspace{20mm}

\vfill
\end{titlepage}

\baselineskip 6.6mm
\renewcommand{\thefootnote}{\arabic{footnote}}
\setcounter{footnote}{0}

\section{Introduction}
Recently, much attention has been focused on the new idea suggested by
Verlinde~\cite{verlinde} in which gravity can be explained as an
emergent phenomenon originated from the statistical properties of
unknown microstructure of spacetime. The essential part of this
idea is based on two key ingredients: 
holographic principle and equipartition rule of the energy. With help
of these principles, the Newton's law of gravity was derived by
interpreting it as an entropic force i.e., force on a test particle at
some point was defined as the product of the entropy gradient and the
temperature at that point, and relativistic generalization leads to
the Einstein equations.
This entropic formulation of gravity has been used to study
thermodynamics at the apparent horizon of the
Friedmann-Robertson-Walker universe~\cite{Shu:2010nv}, Friedmann
equations~\cite{Cai:2010hk,Pad:10013380},
entropic correction to Newtonian gravity ~\cite{Smolin:2010kk, PNicolini, LModesto},
holographic dark
energy~\cite{Li:2010cj, efs, danielson, YFCai}.  There have been many works
for the entropic force in cosmological models~\cite{Gao:2010fw, zgz,
  wang:y, Wei:2010ww} and the black hole
backgrounds~\cite{Myung:2010jv,mk,Liu:2010na, Cai:2010sz, Tian:2010uy,
  Kuang:2010gs, EEKL, RAKonoplya, FCaravelli}.

In a spacetime admitting timelike killing vector one can define a
gravitational potential, and the holographic screen is given by
equipotential surface.  In general, the holographic screen can have
multiple disconnected parts depending upon the matter distribution.
The temperature on the holographic screen is given by Unruh-Verlinde
temperature associated with the proper acceleration of a particle near
the screen.  This prescription works well for spacetime with a single
holographic screen, however, there has been no known work for multiple
holographic screens so far.

On the other hand, the observational evidence for late-time
cosmological acceleration~\cite{Perl:1998, Reiss:1998}
gave much impetus on studying the de Sitter space with black holes.
Since Schwarzschild-de Sitter black hole is asymptotically de Sitter
space, it has cosmological event horizon in addition to black hole
horizon and these horizons can form holographic screens.  In fact, the
potential of Schwarzschild-de Sitter space has two equipotential
surfaces for a given potential value, and the two horizons correspond
to equipotential surfaces.  In this paper we investigate the entropic
formulation in the background geometry of the Schwarzschild-de Sitter
space, which provides a model for multiple holographic screens.

In the Verlinde's formalism, two equipotential holographic screens in
the Schwarzschild-de Sitter space have different temperatures.
Thus the whole system cannot be treated as a thermodynamical system in
equilibrium.  In Ref.~\cite{Bousso:1996au}, Bousso and Hawking set up
a reference point in the radial direction, at which force
vanishes. They have pointed out that this reference point can play a
role of a point at infinity in an asymptotically flat space. Besides,
the temperature at this reference point is zero, and thus no thermal
exchange can occur across this point. This makes the reference point
behave like a thermally insulating wall. Therefore, we can regard the
Schwarzschild-de Sitter space as two thermally independent systems:
the inner system in the black hole side and the outer system in the
cosmological horizon side. Gibbons and Hawking also considered similar
construction in a slightly different context~\cite{GH_desitter}: they
constructed two separated thermal equilibrium systems by introducing a
perfectly reflecting wall in the Schwarzschild-de Sitter space for the
calculation of the Hawking temperatures of black hole and cosmological
horizons.

Based on the above consideration, we apply the Verlinde's formalism to
each system.  In the Schwarzschild-de Sitter case we choose the
holographic screen of equipotential surface having spherical
symmetry. With this choice of holographic screen we show that the
thermodynamic relationship $E = 2TS$ holds for each holographic
screen, where $E$, $T$, and $S$ are the quasilocal energy given by
Komar mass, temperature, and entropy, respectively.  We then check
this result with the known cases: i) when the holographic screens lie
at the black hole and cosmological horizons, ii) in the Nariai limit.


In the following section, we briefly review the Verlinde's formalism
of entropic approach to gravitational interaction.  In
section~\ref{sec:sch.dS}, we apply the Verlinde's formalism to a
Schwarzschild-de Sitter space which provides a prototype of multiple
holographic screens.  Finally, we summarize our results.  In this
paper, we adopt the convention $c=k_B=\hbar =1$.

\section{Verlinde's entropic formalism}
\label{sec:setup}
According to the Verlinde's formalism~\cite{verlinde}, gravity is an
entropic force emerging from coarse graining process of information
for a given energy distribution. In this process, information is
stored on holographic screens.  In the nonrelativistic case, the
holographic screens correspond to Newtonian equipotential surfaces and
the holographic direction is given by the gradient of the potential.

In a curved spacetime with a timelike Killing vector $\xi^\mu$, the
generalized Newton's potential is given by
\begin{equation}
  \label{potential}
  \phi = \frac12 \ln (-\xi^\mu \xi_\mu ).
\end{equation}
This potential can be used to define a foliation of space.  For a
particle with a four velocity $u^\mu$, its proper acceleration is
given by $a^\mu = u^\nu \nabla_\nu u^\mu$.  In terms of the potential
$\phi$ and the Killing vector $\xi^\mu$, the velocity and the
acceleration can be written as
\begin{align}
  u^\mu &= e^{-\phi} \xi^\mu, \label{u:killing} \\
  a^\mu &= - \nabla^\mu \phi, \label{a:killing}
\end{align}
where the Killing equation has been used to derive
Eq.~(\ref{a:killing}). In Eq.~(\ref{a:killing}), the acceleration is
normal to holographic screen. The Unruh-Verlinde temperature on
the screen is defined as
\begin{equation}
  \label{T:def}
  T = \frac{1}{2\pi} e^\phi n^\mu \nabla_\mu \phi,
\end{equation}
where $n^\mu$ is the unit outward pointing vector normal to the screen
and to the Killing vector.
The ``outward'' indicates that the potential increases along $n^\mu$,
\textit{i.e.,} the normal vector can be written as
\begin{equation}
  \label{def:n}
  n_\mu = \frac{\nabla_\mu \phi}{\sqrt{\nabla_\nu \phi \nabla^\nu \phi}}.
\end{equation}
In Eq.~(\ref{T:def}), a redshift factor $e^\phi$ is inserted because
the temperature is measured with respect to the reference point.  For
asymptotically flat space this reference point corresponds to spatial
infinity. In the Schwarzschild-de Sitter case, we choose this
reference point as the Bousso-Hawking reference
point~\cite{Bousso:1996au} to be explained in the next section.

We denote the number of bits on the holographic screen $\mathcal{S}$
by $N$ which is assumed to be proportional to the area of the
screen~\cite{verlinde},
\begin{equation}
  \label{N:assume}
  N = \frac{A}{G}.
\end{equation}
Applying the equipartition rule of the energy, each bit of holographic
screen contributes an energy $T/2$ to the system, and the total energy
on the holographic screen can be written as
\begin{equation}
  \label{E:equipartition}
  E = \frac12 \oint_{\mathcal{S}} T dN.
\end{equation}
Note that in the above expression the temperature $T$ on the screen is not
constant in general.  Substituting Eqs.~(\ref{T:def})
and (\ref{N:assume}) into Eq.~(\ref{E:equipartition}), the energy
associated with the holographic screen can
be rewritten as
\begin{equation}
  \label{E:pot.area}
  E = \frac{1}{4\pi G} \oint_{\mathcal{S}} n^\mu \nabla_\mu e^\phi dA.
\end{equation}
This expression is the conserved Komar mass associated with
timelike Killing vector $\xi^\mu$.


\section{Schwarzschild-de Sitter black hole}
\label{sec:sch.dS}

Now, we consider a spherically symmetric Schwarzschild-de Sitter black
hole as a model of multiple holographic screens.  The Schwarzschild-de
Sitter space is described locally by the line element,
\begin{equation}
  \label{metric:ds}
  ds^2 =
 -f(r) dt^2 + \frac{dr^2}{f(r)} + r^2 (d\theta^2 + \sin^2
  \theta \, d\varphi^2),
\end{equation}
with
\begin{equation}
  \label{f}
  f(r) = 1 - \frac{2GM}{r} - \frac13\Lambda r^2,
\end{equation}
where $G$ and $M$ are the gravitational Newton's constant and the
mass parameter, respectively, and the cosmological constant will be
taken as $\Lambda = 3/\ell^2$.  When $0 < M < M_{\mathrm{max}}\equiv
\ell/(3^{2/3} G)$ static region exists between two horizons with radii
$r_b$ and $r_c$, the black hole and cosmological event horizons.  For
$M = M_{\mathrm{max}}$, the two horizons coincide, the Nariai
limit. In the Nariai limit, there exists no timelike Killing vector.

\begin{figure}[pbt]
  \centering
  \includegraphics[width=0.5\textwidth]{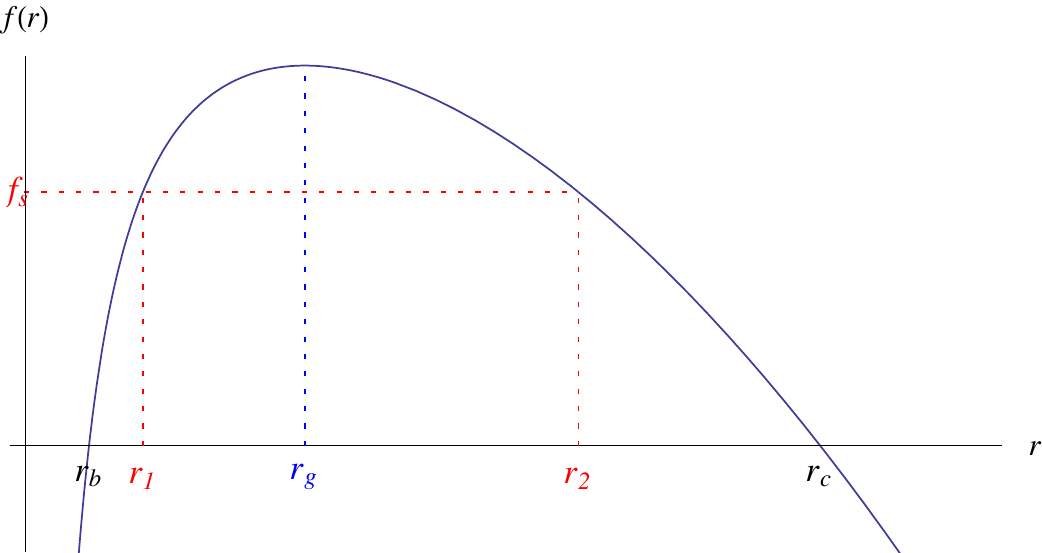}
  \caption{The Schwarzschild-de Sitter space has the event horizon of
    the black hole at $r=r_b$ and the cosmological event horizon at
    $r=r_c$. At $r=r_g$, the proper acceleration vanishes. For a given
    potential value, there exist two screens at $r=r_1$ and $r=r_2$,
    and each screen has different temperature. Note that the unit
    normal vectors on both screens direct to the surface $r=r_g$.}
  \label{fig:screen}
\end{figure}

In order to get the potential of the Schwarzschild-de Sitter spacetime,
we first consider a timelike Killing vector of Eq.~(\ref{metric:ds}),
given by

\begin{equation}
  \label{xi}
  \xi^\mu  = \gamma\left( \partial/\partial t \right)^\mu,
\end{equation}
where $\gamma$ is a normalization constant.  If space is
asymptotically flat, we may choose the standard Killing vector
normalization, $ \gamma = 1$.  Since Schwarzschild-de Sitter space is
not asymptotically flat, we encounter a difficulty in taking the
normalization of Killing vector.  To avoid this, Bousso and
Hawing~\cite{Bousso:1996au} chose a normalization such that the norm
of the Killing vector becomes unity at the region where the force
vanishes, the gravitational attraction is exactly balanced out by the
cosmological repulsion. Adopting this normalization corresponds to
choosing a special observer who follows geodesics.

%
Since the magnitude of the acceleration of a particle in the
Schwarzschild-de Sitter spacetime is obtained as $a = \sqrt{a^\mu
  a_\mu} = |f'(r)|/\sqrt{2f(r)}$, the geodesic point with no
acceleration is given by
\begin{equation}
  \label{def:rg}
  r_g = (GM\ell^2)^{1/3}.
\end{equation}
With this normalization, the gravitational potential is obtained from
Eq.~(\ref{potential}),
\begin{equation}
  \label{potential:f}
  \phi = \frac12 \ln (\gamma^2 f)
= \frac12 \ln  \frac{f(r)}{f(r_g)}.
\end{equation}
For a given potential value $\phi_s$, there exist two equipotential
surfaces at $r=r_1$ and $r=r_2$ as shown in Fig.~\ref{fig:screen}.
Then, the Unruh-Verlinde temperature on each screen is given by
\begin{equation}
  \label{T}
  T = \frac{1}{2\pi} e^\phi n^\mu \nabla_\mu \phi =
  \gamma \frac{|f'(r)|}{4\pi},
\end{equation}
where the unit normal vector $n^\mu$ is given by $n^\mu = \delta^\mu_r
\sqrt{f}$ for $r<r_g$ and $n^\mu = -\delta^\mu_r \sqrt{f} $ for
$r>r_g$.  Note that the temperature of the holographic screen at
$r=r_1$ is different from that of the screen at $r=r_2$.  The
temperature on each screen is given by
\begin{equation}
  \label{T:r}
  T_i = \frac{\gamma}{2\pi} \left| \frac{GM}{r_i^2} - \frac{r_i}{\ell^2} \right|,
\end{equation}
where $i = 1, 2$.

\begin{figure}[tb]
  \centering
  \includegraphics[width=0.5\textwidth]{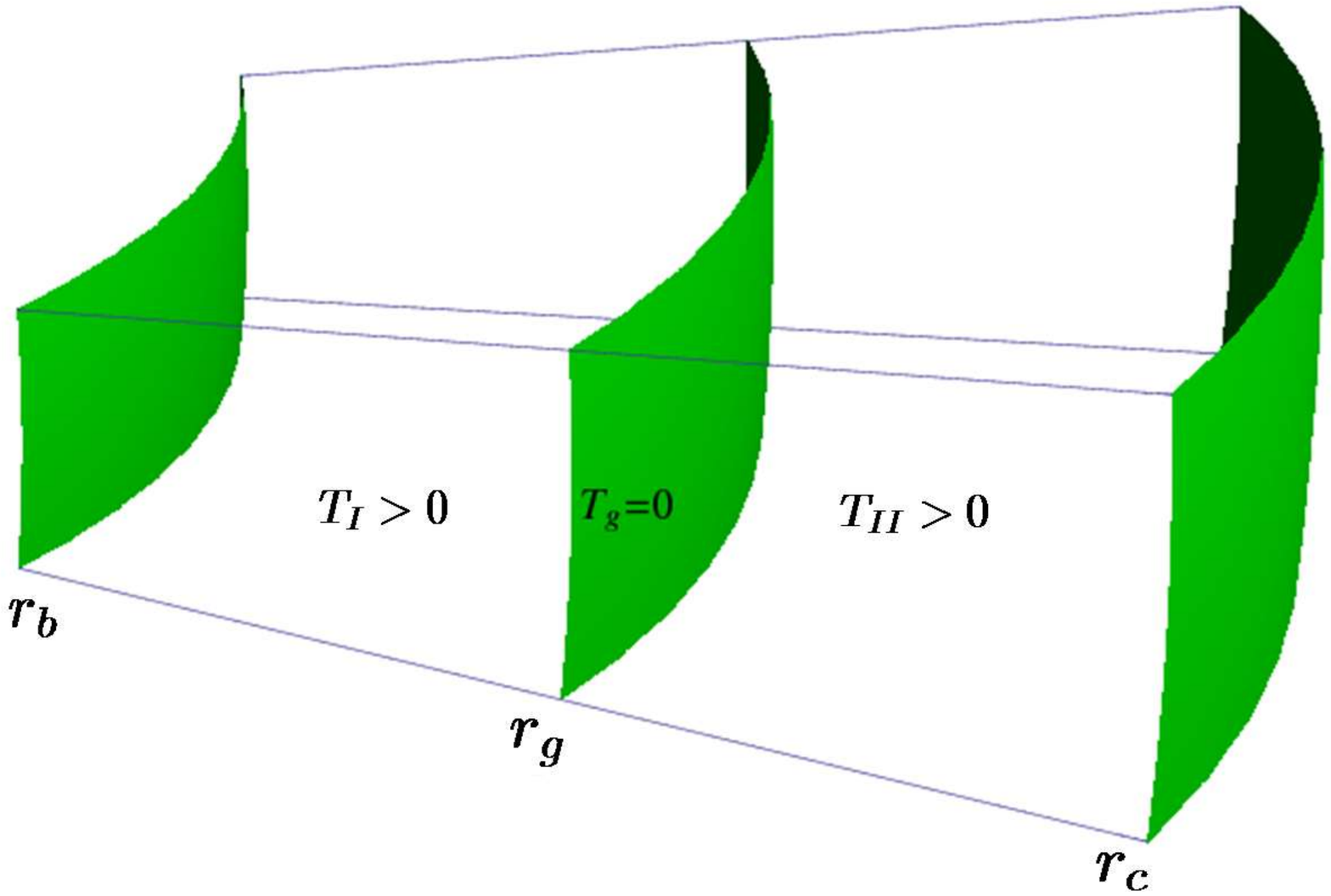}
  \caption{
 We consider the force free reference point of Bousso and Hawking
as a separating boundary dividing the system into two subsystems.
Since the temperature of each subsystem is  above zero
  and the boundary between them is maintained at zero,
thermal exchange does not occur between the two subsystems.}
  \label{fig:tempq}
\end{figure}

The temperature becomes zero at the Bousso-Hawking reference point
$r=r_g$ from Eq.~(\ref{T}). Now, assume that the region between the
black hole and cosmological horizons is separated by a boundary at the
reference point $r=r_g$ as shown in Fig.~\ref{fig:tempq}.
Then the two regions divided by this boundary
cannot have thermal exchange between them because the temperature on
this boundary is kept at zero always in our static geometry
setup. Thus, we can regard this boundary as a thermally insulating
wall.  Therefore, the two regions separated by the surface at $r=r_g$
can be thought as independent systems: the total system becomes the
sum of two independent systems, the inner ($r<r_g$) and outer
($r>r_g$) regions.  The concept of thermally insulating wall in our
consideration is similar to that of perfectly reflecting wall in the
Gibbons-Hawking's work~\cite{GH_desitter}: they constructed two
separated thermal equilibrium systems by introducing a perfectly
reflecting wall in Schwarzschild-de Sitter space for the calculation
of the Hawking temperatures of black hole and cosmological horizons.

\begin{figure}[pbt]
  \centering
  \includegraphics[width=0.5\textwidth]{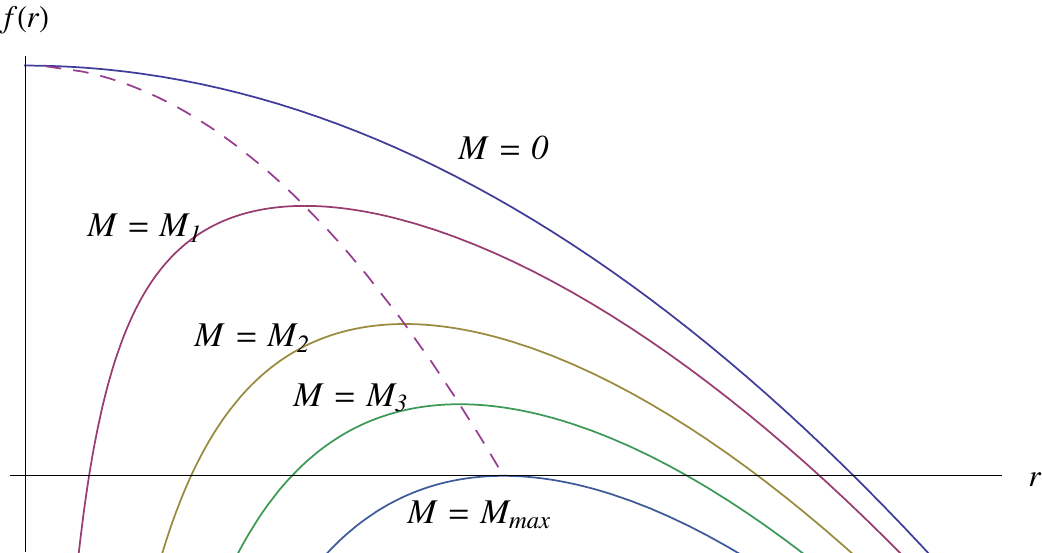}
  \caption{The geodesic point with no acceleration is plotted by the
    dashed line. The metric approaches the pure dS spacetime as the
    mass parameter of Schwarzschild-de Sitter space goes to
    zero. ($0<M_1 < M_2 < M_3 <M_{\mathrm{max}}$)}
  \label{fig:mass}
\end{figure}

\begin{figure}[pbt]
  \centering
  \includegraphics[width=0.5\textwidth]{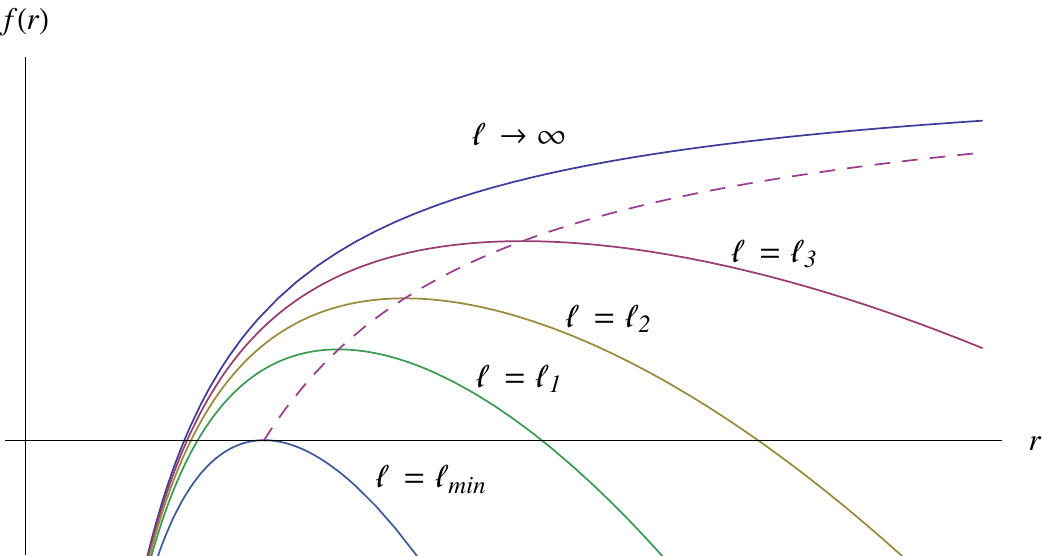}
  \caption{The geodesic point with no acceleration is plotted by the
    dashed line. The metric approaches the Schwarzschild black hole as
    the cosmological constant $\Lambda = 3/\ell^2$ goes to
    zero. ($\ell_{\mathrm{min}} < \ell_1 < \ell_2 < \ell_3 < \infty$)}
  \label{fig:cosmo}
\end{figure}

This can be also understood as follows.  The line
element~(\ref{metric:ds}) approaches the pure de Sitter spacetime when
$M$ goes to zero and the pattern of the metric for $r>r_g$ has a
similarity to that of the pure de Sitter spacetime (see
Fig.~\ref{fig:mass}). And the spacetime approaches the Schwarzschild
black hole with asymptotically flat spacetime when $\Lambda$ goes to
zero and the pattern of the metric for $r<r_g$ has the similarity to
that of the Schwarzschild black hole (see Fig.~\ref{fig:cosmo}). This
suggests that the whole system has the characteristics of both
Schwarzschild black hole and pure de Sitter spacetime.

Plugging the potential~(\ref{potential:f}) into the
energy~(\ref{E:pot.area}) gives the same result from the Komar energy
for the Schwarzschild-de Sitter black hole,
\begin{equation}
  \label{E:g}
  E = \frac{1}{4\pi G} \oint_{\mathcal{S}} \nabla^\mu \xi^\nu
  \sigma_\mu n_\nu dA,
\end{equation}
where $\sigma_\mu$ is the unit normal timelike vector perpendicular to
the hypersurface surrounded by the screen $\mathcal{S}$.
%
Since $\sigma_\mu = - \sqrt{f} \, \delta_\mu^t$, the Komar
energy~(\ref{E:g}) becomes
\begin{equation}
  E_i = \gamma \frac{r_i^2 |f'(r_i)|}{2G} = \gamma \left| M -
    \frac{r_i^3}{G\ell^2} \right|,  \label{E:r}
\end{equation}
for each screen at $r=r_i$ ($i = 1, 2$).

If the associated holographic entropy is given by
\begin{equation}
    S_i = \frac{A_i}{4G} = \frac{\pi r_i^2}{G}, \label{S:r}
\end{equation}
 then  with Eqs.~(\ref{T:r}) and
(\ref{E:r})  the thermodynamic relation $E_i = 2 T_i
S_i$ holds for each system.
This relation certainly holds for event horizons.\footnote{
In Refs.~\cite{Pad:0912,RBan}, it was shown that this relation holds when
the equipartition rule
of energy is assumed for event horizons of stationary spacetimes.}
When the
spacetime is static and spherically symmetric,
we can also get this relation  directly from Eq.~(\ref{E:equipartition})
with the relation \eqref{S:r},
since the temperature on the holographic screen is constant.
Note that the thermodynamic relation $E=2TS$ does not hold for the
whole system,
since the energy and entropy are additive and the temperatures on the holographic screens are different.

Now, we check the validity of our formulation in two specific
cases. First, we consider the case when the holographic inner and
outer screens become the event horizon of black hole and the
cosmological horizon, respectively. As the locations of the
holographic screens, $r_1$ and $r_2$, move to the two roots of $f(r) =
0$, $r_b$ and $r_c$, as shown in Fig.~\ref{fig:screen}, the inner
screen becomes the black hole event horizon and the outer one becomes
the cosmological event horizon.  The temperatures on the screens seen
by an observer located at the Bousso-Hawking reference point are given by
\begin{eqnarray}
  T_{b/c}= \frac{1}{\sqrt{1-(9G^2M^2\Lambda)^{1/3}}}
  \frac{1}{4\pi}\left| \frac{2GM}{r_{b/c}^2} - \frac{2\Lambda r_{b/c}}{3}\right|.
\end{eqnarray}

Since the system is composed of the sum of two independent systems, the total
entropy is given by the sum of the entropies of subsystems,
\begin{equation}
  \label{S:total}
  S = S_1 + S_2.
\end{equation}
In the present case, $S_1$ and $S_2$ correspond to the usual entropy
of the black hole and cosmological horizons, respectively. And thus,
our result agrees with the previously obtained entropy of
Schwarzschild-de Sitter space~\cite{GH_desitter, Kas:1996,SdS_entropy,
  BHlee:jkps,SdS_nar_ent}.

Next, we consider the case when the two event horizons, $r_b$ and
$r_c$, approach each other, the Nariai limit~\cite{nariai_bh}. In this
case, the temperature and the energy on each horizon become
\begin{align}
  T_i &\longrightarrow T^{\rm Nariai} = \frac{\sqrt{3}}{2\pi \ell}, \label{T:Nariai}\\
  E_i &\longrightarrow E^{\rm Nariai} = \sqrt{3} \left(\frac{M^2 \ell}{G}
  \right)^{1/3}. \label{E:Nariai}
\end{align}
In this limit, the entropy of each system becomes
\begin{equation}
 S_i  \longrightarrow\ \frac{\pi r_g^2}{G}.\label{S:i:N}
\end{equation}
The total entropy is the sum of the two subsystems', thus it is twice
of the above given entropy~(\ref{S:i:N}).  This agrees with the
entropy of the Schwarzschild-de Sitter black hole in the Nariai limit
obtained in Refs.~\cite{BHlee:jkps,SdS_nar_ent}.

%

In summary, we apply the Verlinde's entropic formalism of gravity to
the Schwarzschild-de Sitter space as a model of multiple holographic
screens.  Since the Unruh-Verlinde temperature vanishes at the
Bousso-Hawking reference point, we can regard two regions separated by
zero temperature barrier as thermodynamically isolated systems and
thus independently apply the entropic formalism to each region.  We
confirm that the Verlinde's formalism agrees with the conventional
result at least in the following cases; i) when the holographic
screens become event horizons, ii) in the Nariai limit.

\section*{Acknowledgments}
This work was supported by the National Research Foundation (NRF) of
Korea grants funded by the Korean government (MEST)
[R01-2008-000-21026-0 and NRF-2009-0075129 (E.\ C.-Y.\ and K.\ K.),
NRF-2009-351-C00109 (M.\ E.), and NRF-2009-351-C00111 (D.\ L.)].



\end{document}